\documentclass[lettersize,journal]{IEEEtran}
\usepackage{amsmath,amsfonts}
\usepackage{algorithmic}
\usepackage{array}
\usepackage[caption=false,font=normalsize,labelfont=sf,textfont=sf]{subfig}
\usepackage{textcomp}
\usepackage{stfloats}
\usepackage{url}
\usepackage{verbatim}
\usepackage{graphicx}
\usepackage[colorlinks,linkcolor=blue]{hyperref}
\usepackage{float}
\usepackage{newunicodechar}
\usepackage{hyperref}
\usepackage{balance}
\hypersetup{
    colorlinks=true,
    urlcolor=black,  
    citecolor=blue   
}

\usepackage{xcolor}
\usepackage{soul}
\sethlcolor{yellow}

\def\BibTeX{{\rm B\kern-.05em{\sc i\kern-.025em b}\kern-.08em
    T\kern-.1667em\lower.7ex\hbox{E}\kern-.125emX}}
\usepackage{balance}

\begin{document}
\title{Attacking Slicing Network via Side-channel Reinforcement Learning Attack}
\author{Wei Shao\thanks{Corresponding Email: wei.shao@data61.csiro.au}, Chandra Thapa, Rayne Holland, Sarah Ali Siddiqui and Seyit Camtepe 
\thanks{Wei Shao, Chandra Thapa, Rayne Holland, Sarah Ali Siddiqui and Seyit Camtepe are with Data61, Commonwealth Scientific and Industrial Research Organisation, Australia (e-mail: wei.shao@data61.csiro.au; chandra.thapa@data61.csiro.au; rayne.holland@data61.csiro.au; sarah.siddiqui@data61.csiro.au; seyit.camtepe@data61.csiro.au)
}
}


\maketitle

\begin{abstract}
Network slicing in 5G and the future 6G networks will enable the creation of multiple virtualized networks on a shared physical infrastructure. This innovative approach enables the provision of tailored networks to accommodate specific business types or industry users, thus delivering more customized and efficient services. However, the shared memory and cache in network slicing introduce security vulnerabilities that have yet to be fully addressed. In this paper, we introduce a reinforcement learning-based side-channel cache attack framework specifically designed for network slicing environments. Unlike traditional cache attack methods, our framework leverages reinforcement learning to dynamically identify and exploit cache locations storing sensitive information, such as authentication keys and user registration data. We assume that one slice network is compromised and demonstrate how the attacker can induce another shared slice to send registration requests, thereby estimating the cache locations of critical data. By formulating the cache timing channel attack as a reinforcement learning-driven guessing game between the attack slice and the victim slice, our model efficiently explores possible actions to pinpoint memory blocks containing sensitive information. Experimental results showcase the superiority of our approach, achieving a success rate of approximately 95\% to 98\% in accurately identifying the storage locations of sensitive data. This high level of accuracy underscores the potential risks in shared network slicing environments and highlights the need for robust security measures to safeguard against such advanced side-channel attacks.
\end{abstract}


\section{Introduction}
\IEEEPARstart{A}s a key technology for 5G and future communication networks, network slicing has gained widespread attention and recognition from operators and the academic community due to its significant advantages in vertical industry customization, service quality assurance, flexibility, and reliability~\cite{zhang2019overview}. Each network slice operates as an independent and isolated network tailored to specific requirements, enabling efficient resource utilization and customization for diverse services and applications~\cite{afolabi2018network}. Network slicing is a crucial facilitator for providing on-demand customized 5G network services alongside enhanced mobile broadband services, utilizing a shared physical network infrastructure~\cite{zhang2019overview}.

In a sliced network scenario, multiple slices share the same underlying network resources, such as bandwidth, processing capacity, and storage space~\cite{zhang2019overview}. This resource sharing enhances the overall utilization of network resources. Although resources like bandwidth and storage space are shared, network slices are isolated from each other. This means that the activities of one network slice do not directly impact other network slices, ensuring security and isolation among multiple applications or services running on the shared infrastructure. However, side-channel attacks typically involve the monitoring of underlying hardware or shared resources, and these malicious activities are often not entirely preventable through the isolation provided by network slicing. While network slicing offers businesses greater flexibility and control over their network resources, it also introduces new security challenges that must be addressed. To ensure the security of network slices, it is essential to carefully assess the risks associated with their implementation and take appropriate measures to mitigate these risks.


Usually, shared slicing networks share the same physical network infrastructure. However, the very nature of this technology, which involves sharing underlying hardware resources such as cache and memory among slices, introduces inherent security risks because attackers can exploit shared resources to launch side-channel attacks. By observing the patterns of cache usage, an attacker might infer sensitive information about the activities of other slices. For instance, variations in cache access times can reveal details about the operations performed by a neighboring slice, potentially leading to the leakage of sensitive data. Even with the encryption mechanisms, information stored in cache is vulnerable to side-channel attacks that exploit indirect data leakage. Timing attacks measure cache access delays during encryption, while techniques analyze cache hits and misses to infer sensitive information. Differential Power Analysis (DPA) can further expose data by monitoring power consumption during cache usage. Additionally, cold boot attacks may recover encryption keys from residual cache data if not properly cleared. These vulnerabilities emphasize the need for stronger defense mechanisms to protect encrypted data in cache.

This paper primarily focuses on cache side-channel vulnerabilities, especially cache-timing attacks. The shared nature of the cache makes it possible for an attacker to utilize the side-channel approach to speculate on another slice’s data and steal keys or other private information~\cite{su2021survey}. Side-channel attacks bypass traditional access control mechanisms enforced by the operating system and hardware~\cite{zhang2019overview}, presenting significant security threats in practice. Due to their stealthy and potent nature, it is crucial to address the security risks they pose and to actively research effective countermeasures against them.

Most cache timing attacks heavily exploit heuristics and expert knowledge to target vulnerabilities. Evaluating side-channel defenses based solely on known attack sequences is inefficient, making it challenging to uncover novel attack strategies~\cite{cui2022macta}.. To tackle these challenges, we propose a novel framework that leverages Reinforcement Learning (RL) to automatically explore cache-timing vulnerabilities in network slicing environments. This paper demonstrates that a timing-channel attack can be modeled as a guessing game, where the attacker attempts to infer the victim's cache state based on observable side effects, such as execution timing. The attacker 'guesses' which cache lines the victim process has accessed, aiming to evict or manipulate them. In this context, an RL agent can learn optimal strategies through repeated self-play in a controlled environment, utilizing either real hardware or simulation platforms commonly employed in architectural studies. The agent refines its actions based on feedback from the environment, effectively learning cache behaviors and identifying potential vulnerabilities. Our experiments demonstrate that this RL framework can automatically adapt to various cache designs and countermeasures, enabling it to identify cache-timing attacks. It discovers both previously known attack sequences and novel, more efficient strategies that can bypass existing defenses, highlighting the power of the RL approach in this domain.

The following are the main contributions of this paper\footnote{Code will be available to the public upon acceptance of this paper.}:
\begin{itemize}
    \item This study is the first to reveal the potential for side-channel attacks to compromise network slicing. Additionally, we introduce a reinforcement learning-based attack framework to automatically explore cache-timing vulnerabilities in these networks. Our framework uncovers both known and previously undiscovered attack sequences, demonstrating its potential to expose new vulnerabilities that may be overlooked by current techniques.
    \item We validate the effectiveness of our proposed attack framework by conducting a series of simulation experiments using a cache simulator. These experiments illustrate how side-channel attacks can successfully compromise network slicing, demonstrating the real-world impact of our approach.
\end{itemize}

\section{Related works\label{sec:Related works}}
This section will show the literature reviews about the relevant research on side-channel attacks and slice network security, providing a theoretical basis for our proposed side-channel network slicing attack method.

\subsection{Side-channel Attack}
A side-channel attack exploits indirect information leakages, such as timing, power consumption, electromagnetic emissions, sound, or vibration, to infer sensitive data or operations from a system, and it is applicable across various technological and computing domains. Since the 1990s, researchers have successively proposed various attack patterns for side-channel attack techniques~\cite{standaert2010introduction}. Instead of directly attacking the algorithms themselves, side-channel attacks exploit unintended information leakage, or side channels, which include Timing-based attacks, Power Analysis attacks, Electromagnetic Analysis attacks, Cache-Timing attacks, or other attacks based on physical properties~\cite{joy2011side}.

Kocher first proposed a timing-based side-channel attack in 1996~\cite{kocher1996timing}. This method involves analysing the execution time of the target system to carry out attacks and obtain encrypted information. Power Analysis attack~\cite{kocher1998introduction} proposed in 1998 by Kocher leverage fluctuations in power consumption during cryptographic operations. Different parts of encryption operations may lead to distinct power consumption patterns, thereby leaking information about the key. Electromagnetic Analysis attack was proposed in 2001 and is performed by measuring the electromagnetic radiation emitted from a device and conducting signal analysis on it~\cite{gandolfi2001electromagnetic,quisquater2001electromagnetic}. The Cache-Timing attack method, as a widely used and efficient side-channel attack technique, was first introduced by Bernstein in 2005~\cite{bernstein2005cache}. This method leverages observable timing differences in accessing cached and uncached memory (hits/misses) to infer sensitive information; implementation methods include Evict + Time~\cite{osvik2006cache}, Flush + Reload~\cite{yarom2014flush+}, Prime + Probe~\cite{osvik2006cache}, Flush + Flush~\cite{gruss2016flush+}, etc.
\begin{enumerate}
    \item \textbf{Evict + Time.} In the Evict + Time method, the attacker evicts a cache set, runs victim code causing cache misses, and measures access time to the evicted set to reveal the victim's memory access.
    \item \textbf{Flush + Reload.} In the Flush + Reload method, the attacker flushes a shared memory location and observes the victim's access time to detect cache hits, which indicate sensitive data access.
    \item \textbf{Prime + Probe.} In the Prime + Probe method, the attacker primes the cache with known data, executes victim code, and measures cache access times to infer information.
    \item \textbf{Flush + Flush.} Unlike the three aforementioned attack patterns, the Flush + Flush side-channel attack technique does not require accessing memory throughout the entire attack process. Because the Flush + Flush attack method causes little or no cache misses, it is more stealthy~\cite{bhebe2019cache}.
\end{enumerate}

In recent years, there have been some new research developments in cache-timing attacks, with the emergence of attack strategies that combine with machine learning methods. For example, attacks using replacement latency differences~\cite{cui2022abusing}, and machine learning methods~\cite{perianin2021end,chabanne2021side}.

Traditional attack methods for exploiting cache timing vulnerabilities are primarily manually executed, with low efficiency and poor flexibility. They are limited to a few fixed patterns, including Flush + Reload, Prime + Probe, Evict + Time, etc. There is a need to exploit vulnerabilities and evade existing detection techniques. Reinforcement learning seeks to discover a policy that produces a sequence of actions maximizing cumulative rewards. RL is particularly suited to exploring cache-timing attack vulnerabilities because it models the attack process as a sequential decision-making problem. RL can automatically generate attack sequences by learning from interactions with the system, aiming to maximize the likelihood of a successful exploit. This approach enables the discovery of complex, multi-step attack patterns that may not be easily identified through other machine learning methods.

\subsection{Network Slicing Security}
Network slicing has emerged as a cornerstone technology in next-generation networks, driven by the convergence of Software-Defined Networking (SDN) and Network Function Virtualization (NFV). While network slicing enhances flexibility and efficiency, it also introduces significant security challenges, which are crucial for safeguarding the integrity and security of 5G networks and future architectures that utilize slicing. The work in~\cite{cunha2019network} highlights various security threats specific to network slicing. In such scenarios, multiple network slices inevitably coexist on shared infrastructure, creating vulnerabilities that can be exploited by malicious actors, such as through side-channel attacks. Furthermore, the study in~\cite{zhang2019overview} examines security risks associated with network slicing from the perspective of User Equipment (UE), noting that UEs often access multiple slices simultaneously. This multi-slice access raises the possibility that a compromised UE could serve as a vector for attacks, bridging security breaches from one slice to another.


\section{Preliminary Study\label{sec:Preliminary Study}}
In this section, we will discuss our preliminary study and demonstrate the feasibility of our proposed RL-based side-channel network slicing attack method. We will present theoretical analyses regarding slicing networks, authentication processes, and cache-based side-channel attacks.

\subsection{Network Slicing}
Network slicing represents a transformative innovation in the advancement of 5G and the anticipated 6G networks, designed to address the varied and evolving demands of modern applications and services. Through the use of advanced virtualization and orchestration technologies, network slicing facilitates the creation of multiple isolated, customizable virtual networks on a shared physical infrastructure. Each slice can be configured to meet distinct performance requirements—such as latency, bandwidth, and reliability—tailoring the network for diverse use cases. This approach not only optimizes resource allocation but also improves operational efficiency and service agility, enabling a wide range of innovative applications across multiple industries. As the need for more sophisticated and specialized services continues to grow, network slicing emerges as a foundational mechanism for unlocking the full potential of 5G and 6G networks, ensuring their capacity to adapt dynamically to an ever-changing technological landscape. This capability, combined with the flexibility and customization inherent in 5G network slicing, allows businesses to manage their network resources more effectively, enhance performance, and deliver an elevated user experience across various applications and industry verticals~\cite{whymatters}.

Our side-channel attack approach exploits a fundamental characteristic of network slicing: the sharing of underlying hardware resources, such as memory and cache, among different applications. Despite the isolation and customization offered by network slices, the concurrent use of shared physical components introduces potential vulnerabilities. By carefully monitoring and analyzing the behavior of these shared resources, our technique can infer sensitive information from co-located slices, effectively bypassing the logical isolation that network slicing aims to provide. This exploitation highlights a critical security challenge in 5G/6G networks, emphasizing the need for robust defenses against side-channel attacks in multi-tenant environments.

\subsection{Authentication Process}
In the context of 5G network slicing, the shared infrastructure and complexities inherent in managing multiple slices concurrently pose significant risks for side-channel attacks. These attacks exploit indirect paths, such as shared resources or information leakage from cryptographic implementations, to extract data or infer slice activities. Despite robust security protocols like TLS for secure communication and OAuth for stringent access control, vulnerabilities may persist due to the overlapping operational domains and the intricate nature of inter-slice interactions. The use of Network Slice Specific Authentication and Authorization (NSSAA) offers granular control over slice access, but it also introduces complexities in credential management and authentication sequences that could be manipulated for side-channel attacks. Enhanced monitoring and adaptive security measures are imperative to address potential vulnerabilities, ensuring that the system’s integrity and the confidentiality of each slice are maintained within the shared network environment. This scenario underscores the necessity for ongoing security assessments and updates to mitigate emerging threats in dynamic network environments.

\subsection{Cache-based side-channel attack}
Cache memory plays a critical role in enhancing performance by storing frequently accessed data closer to the CPU. It typically consists of multiple levels, including L1, L2, and L3 caches \cite{cathe}. The hierarchy of cache levels dictates that if data is present in a higher-level cache (e.g., L1 cache), it should also be found in lower-level caches (e.g., L2 and L3 caches). This inclusiveness is defined as:
\begin{equation}
m \in L1 \longrightarrow m \in L2  \longrightarrow m \in L3.
\end{equation}

Therefore, the L3 cache is inherently shared among different CPU cores~\cite{su2021survey}. The sharing nature of cache memory poses security concerns for network slicing. Since cache memory can store sensitive data, its sharing introduces the possibility of information leakage between different slices or tenants. An attacker with access to one slice may be able to exploit cache-based side-channel attacks to infer information about other slices sharing the same cache.

To mitigate these security risks, measures need to be taken to enhance cache isolation between slices. Techniques such as cache partitioning and cache coloring can be employed to minimize cache interference and prevent unauthorized access to sensitive information. Cache partitioning assigns dedicated cache portions to different slices, ensuring that they do not overlap or share cache entries. On the other hand, cache colouring assigns different cache colors to different slices, segregating their data and minimizing the likelihood of information leakage.

In conclusion, while cache memory is an essential resource for improving performance in network slicing, its sharing nature introduces security risks. By implementing cache isolation techniques like cache partitioning and coloring, the potential for information leakage can be mitigated, ensuring the confidentiality and integrity of data within network slices. Because the cache is shared, an attacker can use the side-channel approach to speculate on other people's data.


Currently, the shared cache can be attacked using Prime+Probe as shown in Fig.~\ref{fig:probe} below:
\begin{figure}[!ht]
\centering
\includegraphics[width=.48\textwidth]{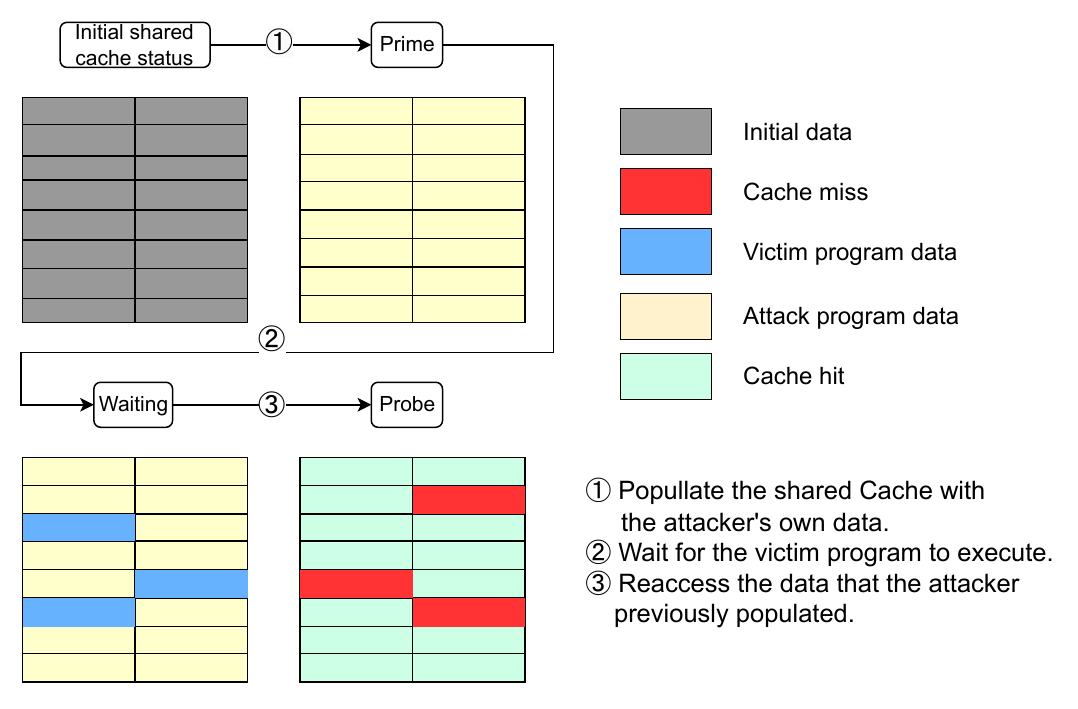}
\caption{Ways to Cache Attacks-Prime+Probe. Different colours indicate different states of the memory block. After steps 1, 2, and 3, the attacker successfully populated the data so that they could re-access them next time.}
\label{fig:probe}
\end{figure}

Step 1: Prime - filling the shared cache with the attacker's own data;

Step 2: Waiting - waiting for the victim (target) slice to execute;

Step 3: Probe - revisit the data previously populated by the attacker; if the access time is longer, it means that the address is used by the victim (the attacker's data in the cache is swiped); if the access time is shorter, it indicates that the victim has not used it (the attacker's data is still in the cache), according to which we can get the victim's use of the address.

In this paper, it is the application of RL that accomplishes the final guessing of the cache block to greatly improve the efficiency of probe detection, simplifying the guessing of the possible replacement cache block to a guessing game that rewards increased accuracy by constantly changing the actions taken by the attacking framework to maximize the reward.

According to our investigation, existing cache security defense mechanisms primarily rely on partitioning and randomization. However, these two defense mechanisms do not effectively address side-channel attacks. Numerous studies demonstrate that cache side-channel vulnerabilities remain a significant challenge. Section~\ref{sec:Defense mechanisms} showcases our related research, providing evidence for the feasibility of RL-based side-channel attack methods.

Based on this attack, assuming that a network slice has been captured, the attack can be prevented by reducing the shared area and making it more challenging to attack (e.g., by setting up different isolation points to prevent reducing the probability of being attacked).

\subsection{Defense mechanisms against Cache-Timing attacks.\label{sec:Defense mechanisms}}
In the past two decades, the risks associated with sharing resources, such as caches, between network slices have been repeatedly underscored~\cite{genkin2023cachefx}. One notable threat is the cache-timing attack, a form of attack that exploits shared cache resources. In such attacks~\cite{liu2014random}, the attacker observes cache contention to determine whether a specific cache line accessed by a victim process has been used, potentially allowing the attacker to infer sensitive information. Memory encryption is a security technique that involves encrypting the contents of computer memory, providing protection against unauthorized access and memory-based attacks by rendering the data unreadable without proper decryption keys. However, related research has revealed severe side-channel vulnerabilities of AMD’s Secure Encrypted Virtualization (SEV)~\cite{li2019exploiting} and Intel’s Software Guard eXtensions (SGX)~\cite{unterluggauer2019meas,wilkeinjection}, both widely used memory encryption methods. Unlike the measures taken for memory security, encryption is not the primary defense mechanism in cache design. Various cache designs have been proposed to address contention-based attacks, with existing mitigation strategies primarily relying on partitioning~\cite{domnitser2012non,wang2007new} and randomization~\cite{liu2016newcache,qureshi2018ceaser,qureshi2019new,werner2019scattercache}. However, these two types of countermeasures are not very effective in defending against attacks~\cite{song2021randomized}.

\section{Methodology\label{sec:Methodology}}

This section will show the details of our proposed methodologies on how to use an RL-based approach to conduct side-channel attacks on the slicing networks. 

\begin{figure*}[!ht]
\centering
\includegraphics[width=0.9\textwidth]{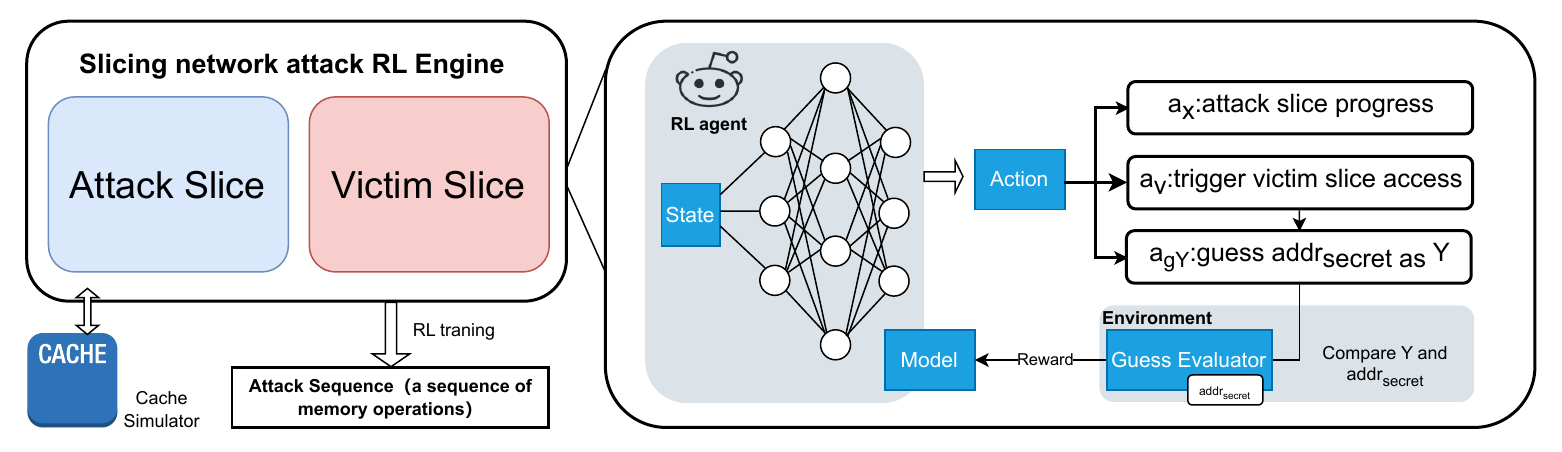}
\caption{RL-based network slicing attack method. The left part of the figure shows the overall framework of the slicing network attack generation method. The right part illustrates how the cache guessing game can be formulated as an RL problem.}
\label{fig:framework}
\end{figure*}

\subsection{Framework Overview}
Fig.~\ref{fig:framework} shows the overall framework of the slicing network attack generation method. The reinforcement learning (RL) engine in the slicing network attack generation method operates based on the target cache implementation, attack slice, victim slice, and RL configuration to generate the attack sequence. Here, we only consider cache implementations based on simulators with certain cache configurations. The attack sequence generated by the RL engine is essentially a sequence of memory operations that human experts can use to carry out attack analysis work to identify potential new attacks. At the same time, the attack sequence generated by the RL engine can also be used to cache the attack demonstration on the simulator. Through this research, we can perform additional manual analysis and classification of attack sequences discovered and generated by the RL engine and apply slicing network attack sequences to multiple cache simulators, achieving a real-world transition to a certain extent.

\subsection{Guessing Game Formulation}
\subsubsection{RL problem}
The right part of Fig.~\ref{fig:framework} illustrates how the cache guessing game can be formulated as a reinforcement learning (RL) problem. In this formulation, the RL engine primarily explores potential slicing network attack strategies by controlling the attack slices. For simplicity, the slicing network attack generation method also allows the RL agent to manage when victim slices are executed. The RL agent's decisions are governed by a policy parameterized by a deep neural network. The environment consists of the secret address ($address_{secret}$) and a guess evaluator that verifies the correctness of each guess. Additionally, the environment interacts with a cache simulator and manages memory access between the attack slice and the victim slice. In the following, we explain and provide an initial formulation of the fundamental elements of the RL problem.

Formally, the cache guessing game can be described using a tuple $\left \{ \mathcal{T},\mathcal{S},\mathcal{P},\mathcal{A},\mathcal{O},\mathcal{R},\gamma \right \}$, where $\mathcal{T}$ is the episode length, $\mathcal{S}$ is the state space, $\mathcal{P}$ is the state transition probability. $\mathcal{A}$ is the action space. $\mathcal{O}$ is the observation space of attack slice, and $\mathcal{R}$ is the reward function for attacker. Lastly, $\gamma \in \left [ 0,1\right ]$ is a reward discount factor.

\subsubsection{Episode}
At the beginning of each episode, the environment randomly initializes the add\_{secret} of the victim slice. The RL agent takes different actions by controlling the attack slice to explore possible attack strategies and obtain corresponding observation results from the environment. Once the RL agent stops executing other actions and decides to guess the $\textup{addr}_{\textup{secret}}$ of the victim slice, the current episode will terminate. Based on the number of steps in the early execution of actions in the current episode and the guess results fed back by the guess evaluator, a certain reward value is given.

\subsubsection{Rewards}
The purpose of the slicing network attack generation method is to guide the RL agent to learn how to correctly find the $\textup{addr}_{\textup{secret}}$. Therefore, when the RL agent guesses the result correctly, the environment will give back a positive reward and vice versa to give feedback a negative reward. In addition, in order to minimize the number of attack steps and achieve optimization of the attack sequence, we set up the environment to give the agent a small penalty every time it takes a step.

\subsubsection{Actions}
RL agent mainly explores slice attack strategies through three actions.

$a_X$: Access memory address $X$. The RL agent can access the cache line at address $X$, which is accessible to the attack slice, and observe whether there is a hit.

$a_v$: Trigger victim slice. The RL agent can trigger the victim slice and induce the victim slice to perform ``secret" access. When the victim slice performs $\textup{addr}_{\textup{secret}}$ access, it may cause the cache status to change.

${a}_{gY}$: Guess the $\textup{addr}_{\textup{secret}}$ value is $Y$, where $Y$ is the address that the victim slice can access. If the guess is about a single secret value, then the current episode ends. Regardless of whether the guess is correct or not, the entire trial phase ends, and then a new trial phase begins; when the RL agent guesses multiple possible secret values, if the guess is incorrect, the trial phase does not end immediately. Instead, the RL agent will change the victim slice's $\textup{addr}_{\textup{secret}}$ guess, allowing the RL agent to continue trying other guesses within the same trial phase. 

\subsubsection{Observations}
When the RL agent takes action $a_X$, the cache simulator will perform the corresponding memory operation, that is, look up the address $X$ in the cache and update the cache status while returning the access delay. When the RL agent takes action $a_v$, the victim slice is triggered, and the cache implementation accesses the address initialized in the current episode. With such an environment set up, the RL agent can explore or generate attack strategies through a series of actions that can be taken.

\subsection{Slicing network attack environment}
\begin{figure}[H]
\centering
\includegraphics[width=.5\textwidth]{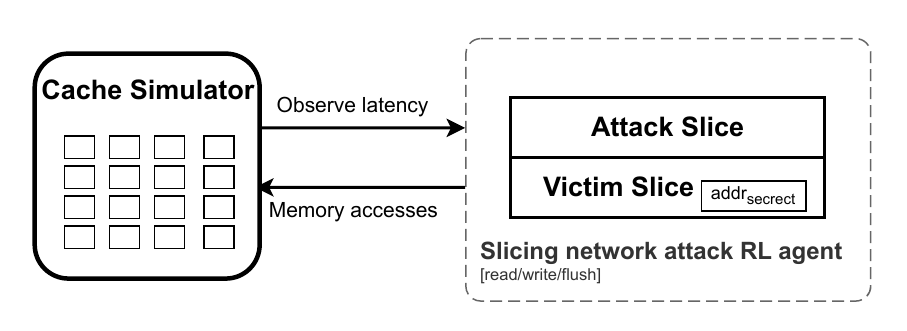}
\caption{Environment for RL-based slicing
network attack.The attack slice
can execute cache read/write/flush operations and see the latency of its own accesses}
\label{fig:scenario}
\end{figure}
Fig.~\ref{fig:scenario} shows the environment for an RL-based slicing network attack. We designed the victim and attack slices to share the same cache in the attack scenario. Furthermore, the attacking slice can observe the latency inherent in its own access operations. In our approach to Cache-Timing attacks, the reinforcement learning (RL) agent estimates the confidential address within the targeted cache slice. This is achieved by manipulating the attacking slice and monitoring variations in cache access latency.

\textbf{Victim Slice.} Research indicates that instances where a victim slice retrieves data from memory locations contingent on confidential information are prevalent. This pattern of secret-dependent memory access is frequently observed in practical applications (e.g., HTTP parsers), widely-used libraries (e.g., OpenSSL), and in the Linux kernel~\cite{johannesmeyer2022kasper,qi2021spectaint,oleksenko2020specfuzz}. 

\textbf{Attack Slice.} The objective of an attack slice is to discern the confidential memory address of the target slice by analyzing the latency patterns in memory accesses. By adopting a strategic approach to choose specific memory addresses for access, an attacker can observe and measure the time taken for these accesses. Moreover, the attacker can initiate the execution of the victim slice and subsequently regain control post-execution. This control enables the attacker to make an educated guess about the secret memory address of the victim slice, provided they have sufficient confidence in their inference. However, it is crucial to note that the attacker's visibility is restricted exclusively to the latency of their memory accesses.

\textbf{RL Action Space.} In the realm of reinforcement learning, the agent is endowed with the capability to command the attack slice to execute one of three distinct actions, denoted as $a_X$, $a_v$, and ${a}_{gY}$. Furthermore, the notation ${a}_{gE}$ represents the scenario in which the agent postulates that the victim slice refrains from any access subsequent to its activation. When the attack slice initiates a guess, two distinct outcomes are conceivable. The first involves the attacker hypothesizing that the target address is $Y$, corresponding to action ${a}_{gY}$. The other scenario entails the presumption that the victim slice will abstain from executing any access operations subsequent to its activation, aligning with action ${a}_{gE}$. The reinforcement learning (RL) agent is programmed to perform one of several actions at each decision point in its operation. These actions are categorized as follows:

\begin{enumerate}
    \item \textbf{Access or Flush Operations}: The agent can either access (denoted as \(a_X\)) or flush (represented by \(a_{fX}\)) data. This choice depends on the current strategy and state of the system.
    
    \item \textbf{Victim Slice Activation}: Another key action is the triggering of the victim slice, symbolized by \(a_v\). This action is integral to the agent's interaction with the system it is navigating.
    
    \item \textbf{Guessing Mechanism}: Lastly, the agent can make guesses, which are represented by either \(a_{gY}\) or \(a_{gE}\). The specific type of guess is contingent on the configuration of the attack and the victim slice.
\end{enumerate}

Each action plays a crucial role in the agent's overall strategy and decision-making process within its operational environment.

\textbf{RL State Space.} The attack slice can discern cache access latency, determining whether it's a hit or miss. To optimize the RL agent's learning, the state is defined by combining historical data of actions and observations. The state space, denoted as $S$, is a composite of various subspace: access latency, executed action, the current step, and the status of the victim slice activation at any given step. This is mathematically represented as $S=\textstyle\prod_{i=1}^{W}\left ( {S}_{lat}^{i}\times{S}_{act}^{i}\times{S}_{step}^{i}\times{S}_{trig}^{i}\right )$, where $W$ is the size of the history window in the observation space. Each subspace has its own definition:
\begin{enumerate}
    \item \textbf{Access Latency Subspace (\( {S}_{lat}^{i} \))}: It represents the cache access state as hit, miss, or not applicable (N.A.).
    \item \textbf{Action Subspace (\( {S}_{act}^{i} \))}: It indicates the specific action taken at step \( i \), chosen from a set of possible actions.
    \item \textbf{Step Subspace (\( {S}_{step}^{i} \))}: It details the particular step in the process.
    \item \textbf{Victim Slice Triggering Subspace (\( {S}_{trig}^{i} \))}: It denotes whether the victim slice has been activated or not at step \( i \).
\end{enumerate}

The RL agent utilizes this comprehensive and structured information to predict the outcome following a secret-dependent memory access by the victim slice.


\textbf{Reward Configuration.} In the context of reinforcement learning (RL), when the agent proposes a guess, the outcome of this guess—correct or otherwise—determines the nature of the response from the environment, which is manifested as either a positive or negative reinforcement. This system is designed to motivate the RL agent towards efficiency and efficacy. Specifically, the environment imposes a negative reinforcement for each action executed by the agent. This approach is strategically employed to incentivise the agent to identify the most succinct and effective sequences for attack. When the cumulative reward within a given episode reaches a threshold of positive value, a deterministic replay method is then utilized to extract and analyses the attack sequences meticulously. The function that delineates our reward design is established as follows
\begin{equation}
\mathcal{R}(S_t, A_t) = 
\begin{cases}
1, & \text{if } a_{gY} = \text{address}_{\text{secret}} \\
-1, & \text{if } a_{gY} \neq \text{address}_{\text{secret}}
\end{cases} .
\end{equation}

In an effort to motivate the reinforcement learning (RL) agent to identify brief sequences of attacks, the environment systematically allocates a diminutive penalty for each action executed. This punitive measure is quantified as a negative step reward, denoted by \(\mathcal{R}(S_t, A_t) = -0.01\), and is applied consistently across all actions undertaken by the agent. This approach ensures that the agent is incentivised to pursue strategies that culminate in shorter sequences, thereby optimizing its performance within the specified parameters of the task.

\textbf{Algorithm and Model.} In our methodology, the aggressor refines their strategy by utilizing the Proximal Policy Optimization (PPO) algorithm, a choice driven by PPO's proven efficacy in learning policies within a Markov game framework. The selection of PPO is attributed to its consistent performance on par with or surpassing alternative approaches. Moreover, PPO's relative ease of fine-tuning is a significant advantage. This is supported by Shusterman \emph{et al}.~\cite{shusterman2019robust}, who highlight the robustness of PPO in various applications. The attacker's policy networks are equipped with Transformer encoders to address the challenges posed by history-dependent observations. These encoders are adept at assimilating information across extensive temporal spans and large datasets. Additionally, Transformer encoders overcome the limitations of vanishing or exploding gradients commonly encountered in Recurrent Neural Networks (RNNs), a point underscored by Parisotto \emph{et al}.~\cite{parisotto2020stabilizing} in their study on network stabilization.

\section{Experiments\label{sec:Experiments}}
In this section, we present an extensive array of experiments designed to evaluate the proposed model. Initially, we detail the experimental setup, including the platform specifications, evaluation metrics, environmental conditions, and hyper-parameter configurations. Subsequently, we discuss three sets of experimental results: training accuracy, training episode length, and the model's performance across varying numbers of cache lines.
\subsection{Experimental Settings}
\subsubsection{Scenario}
We assume there are two slicing networks named \emph{attack network} and \emph{victim network}. The attack network is the compromised slicing network, and the victim network is the shared slicing network that is aimed to be attacked. The attack and victim networks share the same cache, and both can access the cache anytime. The victim network will access the sensitive information in the cache when the shared slicing network sends the requests.
\subsubsection{Hardware}
Experiments were conducted on a machine with an AMD EPYC 7513 32-core processor, NVIDIA Tesla A100 GPU (80GB), and 1TB of RAM.
The AMD EPYC 7513 32-core processor, known for its energy efficiency and scalability, combined with the NVIDIA Tesla A100 GPU, which excels in parallel processing and AI-driven tasks, creates a formidable platform for 5G core networks and future 6G infrastructure. This combination can effectively handle advanced network functions, such as network slicing and real-time data analysis, ensuring robust, high-performance operations. The synergy between these technologies is ideal for Communication Service Providers looking to enhance and future-proof their core network capabilities, as demonstrated by AMD's advancements in 5G technology~\cite{amd2024rewriting}.


\subsubsection{Software Environment} 
To implement our experiment, we utilise the work named AutoCAT~\cite{luo2023autocat}, developed by Facebook Research, which incorporates a specialised cache simulator essential for exploring cache-timing attacks. This simulator forms the backbone of their research, enabling the detailed modelling necessary for their reinforcement learning environment. It supports the training and evaluation of reinforcement learning agents, a crucial aspect of their cybersecurity studies.


\subsubsection{Hyperparameters Settings}
In the major experiment, we aimed to guess a specific memory address from a three-layered cache system. The first layer, L1, contains 16 addresses, representing the fastest and most accessible layer. The second layer, L2, expands to hold 64 addresses, offering a broader range but with slightly increased access time. The largest layer, L3, encompasses 256 addresses, providing the most extensive range but at the slowest access speed compared to the other layers. We conducted 1000 guessing attempts for the experiment to predict the address location within these layers accurately. This setup is designed to test the efficiency and accuracy of the address prediction mechanism across different cache sizes and complexities.

\subsubsection{Evaluation Metrics}
To assess the effectiveness of our proposed method, we focus on two critical metrics: overall guess accuracy and episode length. We define overall accuracy as the ratio of correct guesses to the total number of guesses, mathematically represented by:
\begin{equation}
\text{Successful Rate}\ =\ \frac{\text{Number\ of\ Correct\ Guesses}}{\text{Total\ Number\ of\ Guesses}}.
\end{equation}
This metric reflects the average effectiveness of our attack framework. Meanwhile, episode length, indicating the number of actions within a single episode, measures the method's efficiency. A higher number of actions per guess not only demands more resources but also elevates the risk of detection. Consequently, a shorter episode length indicates a more efficient and safer approach.

\subsection{Results and Analysis}
In this study, we undertook a series of experiments to demonstrate the efficacy of our newly developed side-channel attack methodology.

\begin{table}[]
\caption{Performance of model with different number of cache lines.}
\label{tab:result}
\begin{tabular}{l|l|l}
\hline
\# Cache Lines & Successful Rate (Random) & Successful Rate (Ours) \\ \hline
8 & 12.5\% & 98\% \\ \hline
16 & 6.3\% & 95\% \\ \hline
24 & 4.2\% & 98\% \\ \hline
32 & 3.1\% & 95\% \\ \hline
40 & 2.5\% & 96\% \\ \hline
\end{tabular}
\end{table}

\begin{figure}[!ht]
\centering
\includegraphics[width=.4\textwidth]{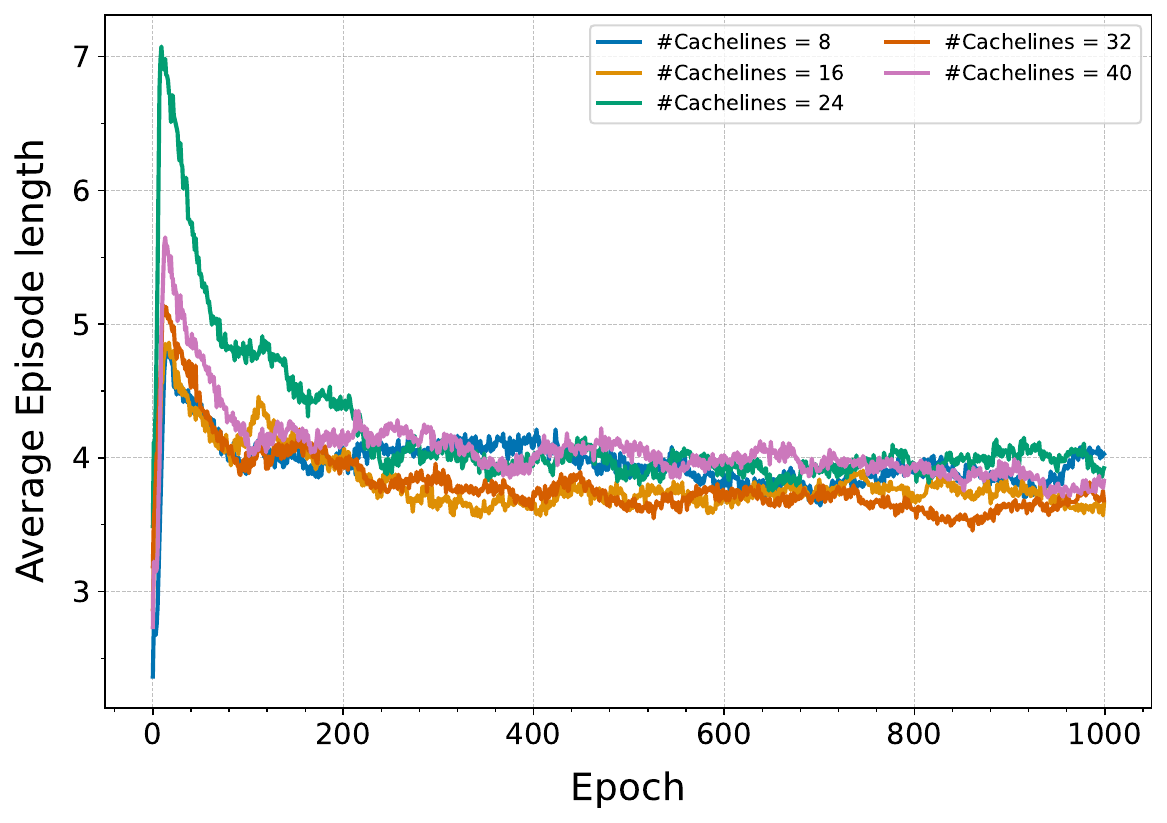}
\caption{Training episode length. The line graph shows the average length of the episodes in every 100 episodes}
\label{fig:eps}
\end{figure}

Table~\ref{tab:result} presents the model's performance across varying cache line sizes, ranging from 8 to 40 in increments of 8. The results indicate a general decrease in success rates for the random guess method as the cache line size increases, starting from the smallest size of 8. Conversely, our proposed framework shows an upward trend in performance with increasing cache line sizes. This pattern suggests that as the number of cache lines grows, identifying the correct cache address becomes increasingly difficult for the random guess method, requiring more sophisticated strategies to accurately pinpoint the correct address. Meanwhile, as previously discussed, encryption is not the central defense strategy in cache design, where speed and efficiency are prioritized~\cite{song2021randomized}. Consequently, once an attacker successfully identifies the correct cache address, the sensitive information stored at that location is vulnerable to theft. These findings underscore the necessity of data security at the cache level. 



Fig.~\ref{fig:eps} illustrates the average duration of episodes during the training process with various number of cache lines. The duration of each episode reflects the number of actions executed in a single round. A shorter episode duration implies that fewer actions are required by the RL agent, leading to reduced resource consumption. Additionally, minimizing the number of actions is crucial, as increased actions heighten the risk of detection by security programs and can introduce delays. Our research findings indicate a significant decrease in the average episode length, from approximately 7 to 4 actions through training. This reduction demonstrates the enhanced efficiency of our agent, capable of accurately guessing the targeted address within an average of 4 actions post-training. The observed consistency in the attack success rate, despite increasing cache line sizes, suggests that the reinforcement learning model has adapted to access patterns that remain effective regardless of cache line dimensions. This may be due to the model's focus on higher-level system behaviors, such as timing discrepancies or cache eviction patterns, which are invariant to cache line size. Additionally, the attack may exploit broader cache set-level patterns, rendering the increase in cache line size ineffective as a countermeasure against this form of side-channel attack.



\section{Conclusion\label{sec:Conclusion}}
This paper proposes an RL-based method to find side-channel vulnerabilities and conduct network slicing attacks automatically. Our automated attack method employs an RL model to explore various cache configurations under different attack and victim slice setups to identify sensitive information. We conducted attack experiments using a cache simulator and successfully generated the attack sequences. Our experimental results demonstrate the feasibility and universality of using RL to explore side-channel attack methods automatically. Our future work will focus on achieving a fully automated attack process, extending from side-channel attacks on network slicing to the authentication process. We aim to design a more universal and efficient RL framework tailored to different cache designs and network slice authentication protocols.

\section{Acknowledgements}
This research work is conducted as part of the 6G Security Research and Development Project, as led by the Commonwealth Scientific and Industrial Research Organisation (CSIRO) through funding appropriated by the Australian Government’s Department of Home Affairs. This paper does not reflect any Australian Government policy position. For more information regarding this Project, please refer to \url{https://research.csiro.au/6gsecurity/}.

\bibliographystyle{ieeetr}
\balance
\bibliography{Mybib}

\end{document}